\documentstyle[preprint,cite,epsf,aps]{revtex}
\begin{document}
\title{\Large Non-Abelian Aharonov-Bohm Scattering of Spin 1/2 Particles\footnote{Copyright by The American Physical Society}}
\author{M. Gomes$^{\,a}$, L. C. Malacarne$^{\,b}$ and A. J. da Silva$^{\,a}$}
  \address{$^{a\,}$Instituto de F\'\i sica, USP\\
 C.P. 66318 - 05315-970, S\~ao Paulo - SP, Brazil.}
\address{$^{b\,}$Departamento de F\'\i sica, Universidade Estadual de
Maring\'a,\\ Av. Colombo, 5790 - 87020-900, Maring\'a - PR,
Brazil.}

\maketitle

\begin{abstract}
  We study the low energy regime of the scattering of two fermionic
  particles carrying isospin 1/2 and interacting through a non-Abelian
  Chern-Simons field. We calculate the one-loop scattering amplitude
  for both the nonrelativistic and also for the relativistic theory.
  In the relativistic case we introduce an intermediate
  cutoff, separating the regions with low and high loop momenta
  integration.  In this procedure purely relativistic
  field theory effects as the vacuum polarization and anomalous
  magnetic moment corrections are automatically incorporated.

\end{abstract}
\section{INTRODUCTION}

Starting from different perspectives, a scalar non-Abelian
Aharonov-Bohm (AB) effect has been discussed by several
authors\cite{Lee,WuYang,Horvathy,Verlinde}.  This subject has
interesting implications to the physics of peculiar objects as cosmic
string and black holes; it has also applications to some aspects of
gravitation in
2+1dimensions\cite{Witten,Alford,Wilczek,Alford2,Kibble,Vilenkin}.
Cosmic strings, for example, may have trapped non-Abelian magnetic
flux tubes so that the scattering of charged particles by these
strings is just a manifestation of the non-Abelian AB effect.

The study of the AB effect was started through the exact calculation
of the scattering amplitude of scalar particles by a thin magnetic
flux tube at the origin\cite{Aharonov}. As it is nowadays well-known,
in that situation the perturbative Born approximation fails to
reproduce the expansion of the exact result\cite{Rafaeli} and,
moreover, the second term of the Born series is divergent. This
discrepancy is due to the fact that the perturbative wave function
does not satisfy the same boundary condition as the exact one.
Actually, in a perturbative treatment for a nonrelativistic field
theory describing spinless Abelian particles scattered through a
Chern-Simons field, it was shown that to eliminate the divergences, to
recuperate the scale invariance and to reproduce the result of the
expansion of the exact solution, it is necessary to add a contact term
$(\phi^* \phi)^2$\cite{Bergman}.

Recently, the perturbative treatment was applied to relativistic
non-Abelian scalar particles\cite{Malacarne} minimally coupled to
non-Abelian CS field.  By considering the low momentum limit, it was
shown that, up to leading order, the same results, got through the
calculation of a non-Abelian nonrelativistic field theory\cite{Bak},
is obtained. In the next to leading approximation, new corrections appear
which are absent in the direct nonrelativistic approach. These
corrections also differ from the ones got in the Abelian
theory\cite{Jorge}.

By analyzing the Abelian AB effect, it has been verified that new
features appear if spin is introduced
\cite{Hagen1,Marcelo1,Marcelo3,Marcelo2}.  For example, the Pauli's
magnetic term plays the role of a contact interaction and no quartic
self-interaction is needed. Besides that, as shown in \cite{Marcelo2}
new effective low momentum interactions are induced if one starts from
a fully relativistic theory.

Completing our study of the non-Abelian AB effect began
in\cite{Malacarne}, in this work we analyze the AB scattering for
non-Abelian spin 1/2 particles.  We start by calculating the AB
scattering in a nonrelativistic setting. We then consider
the AB scattering from a more basic standpoint, starting from a
relativistic quantum field theory, and then taking the appropriate
nonrelativistic limit of the scattering amplitudes. One of the
advantages of such procedure is in the fact that it automatically
incorporates quantum radiative corrections as the vacuum polarization
and induced magnetic moment. To take the nonrelativistic limit most
easily, we use an intermediate auxiliary cutoff separating the low and
high loop momenta in the Feynman integrals.  As it happened in our
previous studies, it is also convenient to work in the Coulomb gauge,
since in this gauge the Chern-Simons propagator depends only on the
spatial part of the loop momentum variable.

\section{NONRELATIVISTIC THEORY}

We consider the non-Abelian Pauli--Schr\"odinger model for fermions
minimally coupled to a non-Abelian Chern-Simons field specified by
the Lagrangian

\begin{eqnarray}
{\cal L} &=& -\Theta \varepsilon^{\alpha\beta\lambda} \mbox{tr}
\left( A_\alpha \partial_\beta A_\lambda +\frac{2 g}{3} A_\alpha
A_\beta A_\lambda \right) + \psi^\dagger \left[ i\partial_t +
\frac{(\bf{\nabla} - g \bf{A})^2}{2m} + i g A_0 - i
\frac{g}{2m}B\right]\psi \nonumber \\ &-&\frac{1}{\xi} \mbox{tr}
(\bf{\nabla}.\bf{A} )^2 - c^{* a} (\delta_{ab}{\bf \nabla}^2 + g 
\varepsilon_{abc}
{\bf A}^c. {\bf \nabla})c^b \; ,
\label{1.1}
\end{eqnarray}

\noindent
where $\psi$ is a one-component anticommuting field, belonging to
the fundamental representation of the $SU(2)$ group, and
$A_\mu=A_\mu^a T^a$, with $T^a$ being the generator of the Lie Algebra
of $SU(2)$ satisfying

\begin{equation}
[T_a,T_b]=  \varepsilon_{abc}T^c\; ,
\label{1.2}\end{equation}

\noindent
and normalized such that

\begin{equation}
T^a T^b= -\frac{\delta_{ab}}{4} I + \frac12 \varepsilon^{abc} T_c 
\label{1.3}
\end{equation}

\noindent
The term containing the ``magnetic'' field $B$, is the Pauli term
and $c$ is the ghost field needed to guarantee unitarity. For convenience,
we will work in a strict Coulomb gauge obtained by letting $\xi\rightarrow 0$. 

We will use a graphical notation where the CS field, the matter
field and the ghost field propagators are represented by wavy,
continuous and dashed lines respectively. The analytic expression
for the $A_\mu$ free propagator is

\begin{equation}
D^{\mu\nu}(k)_{ba} = D^{\mu\nu}(k) \delta_{ba}
=\frac{1}{\Theta}\varepsilon^{\mu\nu\lambda}\frac{\overline{k}_{\lambda}}
{{\bf k}^2}\;\delta_{ba} \;,  \label{1.4}
\end{equation}

\noindent
where $\overline{k}_{\lambda}\equiv (0,\bf {k})$. The matter field
propagator is

\begin{equation}\label{1.6}
  S(p)_{nm}  = S(p) \delta_{nm}
= \frac{i}{p_0-\frac{{\bf p}^2}{2m} +  i\epsilon} \delta_{nm},
\end{equation}

\noindent
whereas the ghost field propagator is

\begin{equation}\label{1.7}
  G(p)_{ba}  = G(p) \delta_{ba} = \frac{-i}{{\bf
  p}^2}\;\delta_{ba}\; .
\end{equation}

\noindent
Since the $B$ field occurs in Eq. (\ref{1.1}) it is convenient to have at
hand

\begin{equation}\label{1.5}
\Delta_B^{ba}(x) = <TB^b(x) A_{0}^{a} (0)> = -\frac{i}{\Theta} \delta^{(3)} (x)
\delta^{ba},
\end{equation}

\noindent
which is the only nonvanishing propagator involving the $B$ field;
graphically it will be represented by a dotted line. Expression
(\ref{1.5} shows that the Pauli term, i . e., the interaction
$\psi^\dagger B \psi$, plays the same role as the quartic term
$(\phi^\dagger \phi)^2$ in the scalar case.

\begin{figure}
\centerline{ \epsfbox{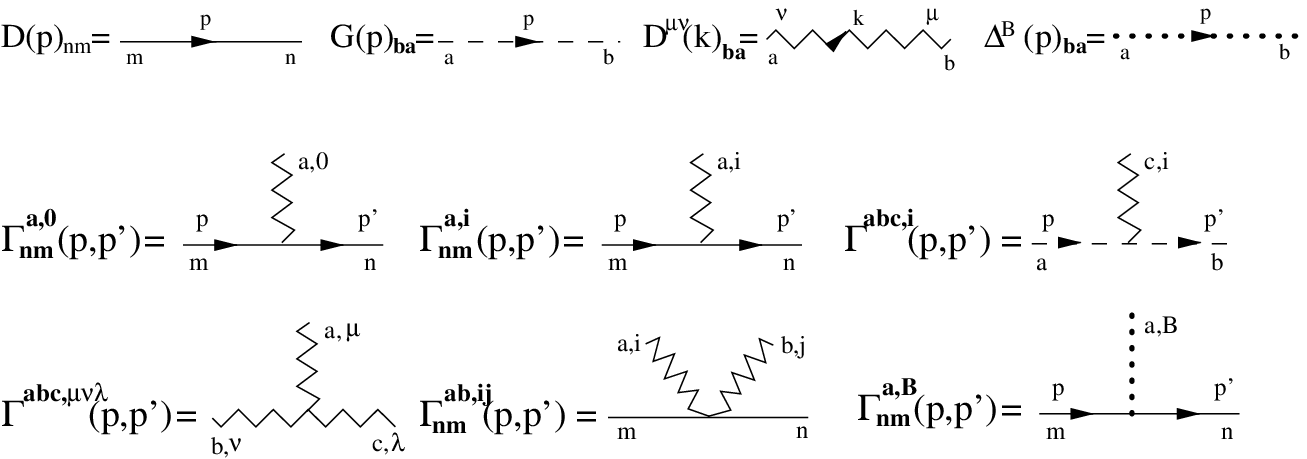}}
\caption{{\it Feynman rules - nonrelativistic theory}} \label{Feynman41}
\end{figure}

The graphical representation for the vertices is given in
Fig. \ref{Feynman41} and the corresponding analytical expression are

\begin{eqnarray}
 & &\Gamma^{a,0}_{nm}(p,p^\prime)= -g(T^{a})_{nm},\\
 & & \Gamma^{a,i}_{nm}(p,p^\prime)=-\frac{g}{2m} (T^a)_{nm}
(p^{i}+p^{\prime i}),\\
 & &\Gamma^{ab,ij}_{nm}(p,p^\prime)= -
i\frac{g^2}{2m} (T^{a} T^b + T^b T^{a})_{nm} g^{ij},\\
 & &\Gamma^{a,B}_{nm}(p,p^\prime) = \frac{g}{2m}(T^a)_{nm},\\
 & &\Gamma^{abc,\mu\nu\lambda}(p,p^\prime) = i  g \Theta
\varepsilon^{abc} \varepsilon ^{\mu\nu\lambda},\\
 & &\Gamma^{abc,i}_{nm}(p,p^\prime)=-g \varepsilon^{abc} p^{\prime i}
\delta_{nm}.
\label{1.8}
\end{eqnarray}

In the  tree approximation and in
the center-of-mass frame the two body scattering amplitude is given by

\begin{equation}
{\cal M}(\theta)= \frac{i g^2}{m\Theta} [T^{a}\otimes T_a] \left[
1 + i\frac{\sin \theta}{(1-\cos \theta)}\right] \; , \label{1.10}
\end{equation}
\noindent
where $\theta$ is the scattering angle. Here and in what follows we employ
a simplified notation where the isospin indices are omitted. If the incoming
and outgoing particles have isospin ($n,m$) and ($n',m'$) the total amplitude
for the process is given by

\begin{equation}\label{1.9}
 {\cal M}_{n'm';nm} = \langle n^\prime,m^\prime \mid {\cal M}(\theta) 
\mid n,m\rangle
- \langle m^\prime,n^\prime \mid {\cal M}(\theta+\pi) \mid
n,m\rangle .\;
\end{equation}

\begin{figure}
\centerline{ \epsfbox{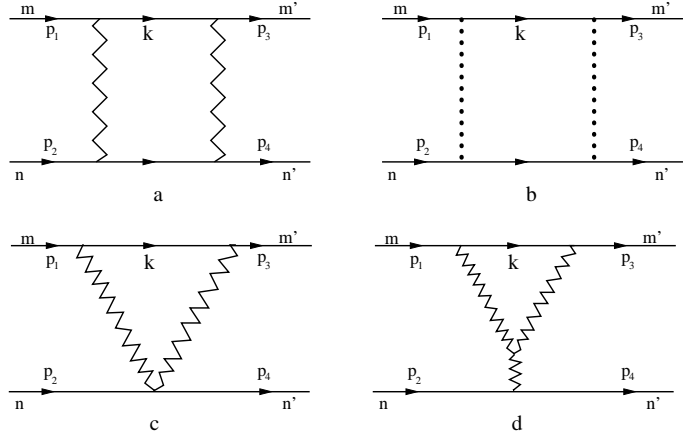}} \caption{{\it  1-loop scattering
- nonrelativistic theory .}} \label{fig4.2}
\end{figure}

The one-loop contribution to AB scattering is depicted in the Fig.
\ref{fig4.2}. The incoming and outgoing fermions are assumed to have
momenta $p_1=({\bf p}_{1}^{2}/2m, {\bf p}_{1})$, $p_2=({\bf
p}_{2}^{2}/2m, {\bf p}_2)$ and $p_{3}=({\bf p}_{3}^{2}/2m, {\bf
p}_{3})$, $p_{4} =({\bf p}_{4}^{2}/2m, {\bf p}_{4})$, respectively.
We work in the center-of-mass frame where ${\bf p}_1=-{\bf p}_2={\bf
p}$, ${\bf p}_3=-{\bf p}_4= {\bf p}^\prime$ and $|{\bf p}| =|{\bf
p}^\prime|$. For the first graph, Fig. \ref{fig4.2}(a), we get

\begin{eqnarray}
{\cal M}_{\mbox{a}} (\theta) = \int \frac{d^3k}{(2\pi)^3} &
&\left[\Gamma^{d,\alpha}(p_1+p_2-k,p_4) S (p_1+p_2 -k)
\Gamma^{c,\nu} (p_2,p_1+p_2-k) \right. \nonumber \\ & &\left.
D^{ac}_{\mu\nu} (k-p1)
 D_{\alpha \beta}^{db} (k-p_3) \Gamma^{b,\beta} (k,p_3) S(k)
\Gamma^{a,\mu} (p_1,k) \right ]\; . \label{1.11}
\end{eqnarray}

\noindent
After performing the $k_0$ integration, this gives

\begin{equation}
{\cal M}_{\mbox{a}} (\theta) = -\frac{4ig^4}{m\Theta^2} [T^b T^{a}
\otimes T_b T_a] \int \frac{d^2{\bf k}}{(2\pi)^2} \frac{1}{{\bf
p}^2 -{\bf k}^2 +i\epsilon} \left[ \frac{ ({\bf p}_1 \wedge {\bf
k})({\bf p}_3 \wedge {\bf k})} {({\bf k}-{\bf p}_1)^2 ({\bf
k}-{\bf p}_3)^2} \right ]. \label{1.12}
\end{equation}

\noindent
As a general rule, whenever dealing with divergent spatial integrals
we will introduce a nonrelativistic cutoff $\Lambda_{NR}$. However, in
Eq. (\ref{1.12}) such regulator is not necessary as the integral is
ultraviolet finite.  The final result is

\begin{equation}
{\cal M}_{\mbox{a}} (\theta) = -\frac{ig^4}{4\pi m\Theta^2} [T^b
T^{a} \otimes T_b T_a] \left\{ \log \left[\frac{{\bf q}^2}{{\bf
p}^2} \right] + i\pi \right \}\; . \label{1.13}
\end{equation}

\noindent
where ${\bf q} = {\bf p_3} - {\bf p_1} $ is the momentum transferred.

The same procedure can be used  to calculate the other graphs in
Fig. \ref{fig4.2}. Here the spatial integrals are logarithmically
divergent and are done after the introduction of the aforementioned
cutoff. Graph \ref{fig4.2}(b) gives

\begin{equation}
{\cal M}_{\mbox{b}} (\theta) = -\frac{g^4}{ m^2\Theta^2} [T^b
T^{a} \otimes T_b T_a]\int \frac{d^3 k}{(2\pi)^3} S(p_1+p_2-k)\,S(k), 
\end{equation}

\noindent
from which we obtain

\begin{equation}
{\cal M}_{\mbox{b}} (\theta) = \frac{ig^4}{4\pi m\Theta^2} [T^b
T^{a} \otimes T_b T_a] \left\{ \log
\left[\frac{\Lambda^2_{NR}}{{\bf p}^2} \right] + i\pi \right \}.
 \label{1.14}
\end{equation}

Similarly, graph \ref{fig4.2}(c) corresponds to

\begin{equation}
{\cal M}_{\mbox{c}} (\theta)= 2 \int \frac{d^3 k}{(2\pi)^3}\Gamma^{cd,ij}
D^{ac}_{0i}(k) D^{db}_{j0}(k+q) \Gamma^{b,0}\Gamma^{a,0}.
\end{equation}

\noindent
The $k^0$ integration is straightforward and gives

\begin{equation}
{\cal M}_{\mbox{c}} (\theta)= \frac{ig^4}{2 m\Theta^2} [(T^a
T^{b}+ T^b T^a )\otimes T_b T_a]\int \frac{d^2 k}{(2\pi)^2}
\frac{{\bf k}\cdot ({\bf k}+ {\bf q})}{{\bf k}^2({\bf k}+ {\bf q})^2}.
\end{equation}

Effectuating the remaining integral produces
 
\begin{equation}
{\cal M}_{\mbox{c}} (\theta)=\frac{ig^4}{4\pi m\Theta^2}
[T^bT^{a}\otimes T_b T_a + \frac12 \varepsilon^{cab} T_c \otimes  T_b
T_a ] \left\{ \log \left[\frac{{\bf q}^2}{\Lambda^2_{NR}} \right]
 \right \} \; .
\label{1.15}
\end{equation}

The last diagram, graph \ref{fig4.2}(d) gives

\begin{eqnarray}
{\cal M}_{\mbox{d}} (\theta)&=& 2 \int \frac{d^3 k}{(2\pi)^3}
\left [\Gamma^{b,\nu}(p_2,p_4) D^{ab}_{\nu\mu}(q) 
\Gamma^{ac^\prime d^\prime, \mu\rho\sigma} 
D^{d^\prime d}_{\sigma\alpha}(k-p_3)\nonumber\right . \\
&\phantom a & \left .\Gamma^{d,\alpha} (k,p_3)S(k) \Gamma^{c,\beta}(p_1,k) 
D^{cc^\prime}_{\beta\rho}(k-p_1)\right]
\end{eqnarray}

\noindent
so that, after the $k^0$ integration,

\begin{equation}
{\cal M}_{\mbox{d}} (\theta)= \frac{ig^4}{ m\Theta^2} [\varepsilon^{cab}T_c
\otimes T_b T_a]\int \frac{d^2 k}{(2\pi)^2}\frac{[{\bf q }\wedge {\bf k }-
{\bf p }_1\wedge{\bf p }_3]({\bf q }\wedge{\bf k })}{{\bf q }^2({\bf k }-
{\bf p }_1)^2({\bf k }-{\bf p }_3)^2}
\end{equation}

\noindent
leading to
 
\begin{equation}
{\cal M}_{\mbox{d}} (\theta)=\frac{ig^4}{8\pi m\Theta^2}
 [ \varepsilon^{cab}T_c\otimes T_b T_a]
\left\{ 1-\log \left[\frac{{\bf q}^2}{\Lambda^2_{NR}} \right]
 \right \} \; .
\label{1.16}
\end{equation}

Thus, the  sum of the one-loop contribution is
 \begin{equation}
{\cal M}_{1-loop} (\theta)=\frac{ig^4}{8\pi m\Theta^2}
 [ \varepsilon^{cab}T_c\otimes T_b T_a]
= -\frac{ig^4}{8\pi m\Theta^2}[T^{a} \otimes T_a] \; .
\label{1.17}
\end{equation}

It happens that the nonvanishing result in the last equation is only
due to the regularization used.  Really, as the original expression
was logarithmically divergent, different regularization schemes will
produce results that for the finite part will differ at most by a
constant. This remark holds even for the sum of the Feynman integrals
which is only conditionally convergent and leads to different results
depending on the way it is treated.  In particular, had we used the
dimensional regularization, as it was done in the
reference\cite{Bergman} for the scalar case, Eq.  (\ref{1.17}) would
be zero. Our constant term in that result may be eliminated through a
redefinition of the cutoff $\Lambda_{NR}$ in the Eq.  (\ref{1.16}) or
by adding a counterterm of the form $(\psi^\dagger T^a
\psi)^2$ to the original Lagrangian. In the relativistic theory
the divergences are milder, the graphs are individually
finite and no such counterterms are needed.

\section{RELATIVISTIC THEORY}

\begin{figure}
\centerline{ \epsfbox{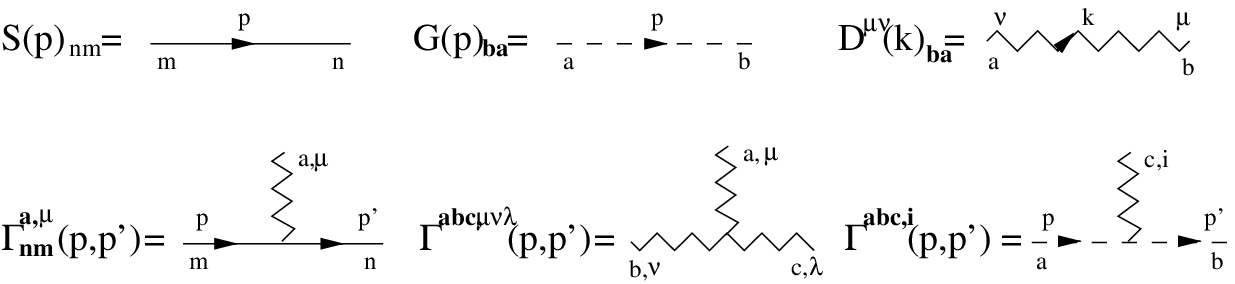}} \caption{{\it Feynman rules
- relativistic theory .}} \label{Feynman42}
\end{figure}

We will now consider the non-Abelian scattering within the full
relativistic context.  The Lagrangian describing the model is

\begin{eqnarray}
{\cal L} &=& -\Theta \varepsilon^{\alpha\beta\lambda} \mbox{tr}
\left( A_\alpha \partial_\beta A_\lambda +\frac{2 g}{3} A_\alpha
A_\beta A_\lambda \right) + i\bar{\Psi} (\not \!\! D -m )\Psi
\nonumber \\ &-&\frac{1}{\xi} \mbox{tr} (\bf{\nabla}.\bf{A} )^2
- c^{* a} ({\bf \nabla}^2 + g \varepsilon_{abc} {\bf A}^c. {\bf \nabla})
c^b \; . \label{4.22}
\end{eqnarray}

\noindent
where $D_\mu=\partial_\mu + g A_\mu$ and $\Psi$ is a two-component
Dirac field belonging to the fundamental representation of the SU(2)
gauge group.  $\psi$ represents
particles and anti-particles with the same spin and we take $m$ to be 
positive so that.  Our graphical
notation is specified in Fig \ref{Feynman42}. The corresponding
analytical expressions for the gauge and ghost field propagators are
the same as in the previous section. The matter field propagator and
the vertices, however, are now given by

\begin{eqnarray}
& &S(p)_{nm} = S(p) \delta_{nm} = \frac{i(\not \! p +m)}{p^2 -m^2
+ i\epsilon} \delta_{nm},\\
 & &\Gamma^{a,\mu}_{nm}(p,p^\prime)= -g (T^a)_{nm}
(\gamma^\mu),\\
 & &\Gamma^{abc,\mu\nu\lambda}(p,p^\prime)= i 
g \Theta f^{abc} \varepsilon ^{\mu\nu\lambda},\\ & &
\Gamma^{abc,i}_{nm}(p,p^\prime) =-g \varepsilon^{abc} p^{\prime
i}\delta_{nm}\;. \label{4.26}
\end{eqnarray}

\begin{figure}
\centerline{ \epsfbox{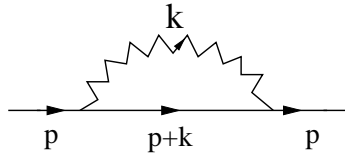}} \caption{{\it Matter field 
self-energy .}} \label{fig4.3}
\end{figure}

\noindent
The model is renormalizable. Actually, without the matter field it was
found that there are no radiative corrections to the Green functions
\cite{Ferrari}.  We can therefore restrict our study of one-loop
renormalization to superficially divergent graphs arising from the
coupling to the matter field, i. e., the 1-loop correction to the
self-energy, vacuum polarization and vertex corrections.  The
nonvanishing self-energy graph depicted in Fig. \ref{fig4.3} is given
by \cite{notacao}

\begin{eqnarray}
\Sigma (p) &=& - \int \frac{d^3k}{(2\pi)^3} \left[\Gamma^{a,\mu}
(p+k,p)
  S(p+k) \Gamma^{b,\nu} (p,p+k) D_{\nu\mu}^{ab} (k) \right]\; 
\nonumber \\ &=& \frac{ig^2}{\Theta} [T^{a} T_a] \int
\frac{d^3k}{(2\pi)^3} \frac{[\gamma^\mu ( \not \!p +\not \! k + m)
\gamma^\nu]\,\, \varepsilon_{\mu\nu\lambda}
\overline{k}^\lambda}{[(p+k)^2 -m^2 +
  i\epsilon]\,\, {\bf k}^2},
\label{4.27}
\end{eqnarray}

\noindent
so that the inverse of the complete fermion propagator is written as
${\cal S}^{-1} (p) = \not \! p - m + i \Sigma $. Notice that the
self-energy is diagonal in isospin space.  After doing the $k_0$
integration we obtain

\begin{equation}
\Sigma (p) = -\frac{ig^2}{8\pi \Theta} [T^{a} T_a]
\int_0^{\Lambda_{0}^{2}} d{\bf k}^2 \frac{1}{w_k} \left\{\frac{m}{{\bf
p}^2} {\bf \gamma} \cdot {\bf p} [1- \epsilon({\bf k}^2 -{\bf
p}^2)] + [1 + \epsilon({\bf k}^2 -{\bf p}^2)] \right\}\; ,
\label{4.29}
\end{equation}

\noindent
where $\epsilon(x)$ is the signal function, $w_k= \sqrt{{\bf k}^2
+m^2}$ and a cutoff $\Lambda_0$ was introduced to take care of the
ultraviolet divergence of the integral.  The integral is easily done
and gives

\begin{equation}
\Sigma(p)= -\frac{ig^2}{2\pi \Theta} [T^{a} T_a]\left [ {\bf\gamma}\cdot
{\bf p} \frac{m}{{\bf p}^2}(w_p -m) + \sqrt{\Lambda_{0}^2+ m^2} - w_p
\right ]
\end{equation}

\noindent
and so, for $\Lambda_0 \rightarrow \infty $,
\begin{equation}
\Sigma(p)= -\frac{ig^2}{2\pi \Theta}
[T^{a} T_a] \left\{ \frac{m \; {\bf \gamma} \cdot {\bf p}-{\bf p}^2 }
{w_p  + m} - m+ \Lambda_0 \right\}\; . \label{4.32}
\end{equation}

\noindent
The linear ultraviolet divergence may be eliminated through the
imposition of an adequate renormalization condition. Due to our use of
the Coulomb gauge, a convenient condition is the one adopted in the
work \cite{Adkins}; denoting the renormalized propagator by ${\cal
S}_R$, this condition reads $ {\cal S}_R(p_0,{\bf p}=0) = S (p_0,{\bf
p}=0)$. Proceeding in this way, we get for the renormalized propagator

\begin{equation}
{\cal S}_R(p) =i\frac{(\not \! p + m) +\alpha (m-w_p) \left[ 1+
\frac{m}{{\bf p}^2} {\bf \gamma}. {\bf p} \right]}{ (p^2-m^2)}\; .
\label{4.41}
\end{equation}

\noindent
where $\alpha = -g^2[T^aT_a]/(2\pi\Theta)$.

\begin{figure}
\centerline{ \epsfbox{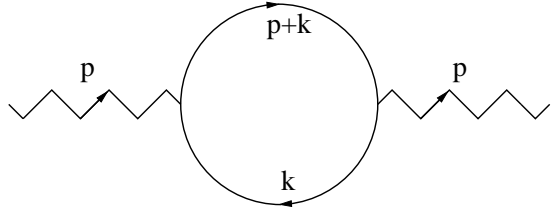}} \caption{\it {Vaccum
polarization}} \label{fig4.4}
\end{figure}

Let us now turn our attention to the vacuum polarization correction.
The only graph that contributes is the one drawn in
Fig. \ref{fig4.4}. As this graph is gauge independent, the would be
linear divergence may be eliminated if one employs a gauge invariant
regularization scheme. Use of dimensional regularization gives

\begin{equation}
\Pi^{\mu\nu}(q)= \frac{ig^2}{4\pi} \mbox{tr} [T^{a}T^b] \left[
\left(g^{\mu\nu} -\frac{q^\mu q^\nu}{q^2}\right) \Pi_1 (q^2) + i m
\varepsilon^{\mu\nu\lambda} q_\lambda \Pi_2 (q^2) \right]\; ,
\label{4.43}
\end{equation}

\noindent
with
\begin{eqnarray}
\Pi_1 (q^2) &=&\int_0^1 dx \frac{2 q^2 x(1-x)}{[m^2 - q^2
x(1-x)]^{1/2}} \approx \frac{q^2}{3|m|} \label{4.44}\\ \Pi_2 (q^2)
&=& \int_0^1 dx \frac{1}{[m^2 - q^2 x(1-x)]^{1/2}} \approx
\frac{1}{|m|}\; , \label{4.45}
\end{eqnarray}

\noindent
where the expressions on the right of these equations are the leading
approximations for low momenta $q$. From these results, we see that for
low momentum a Yang-Mills term may be induced, as one would expect
on general grounds.

\begin{figure}
\centerline{ \epsfbox{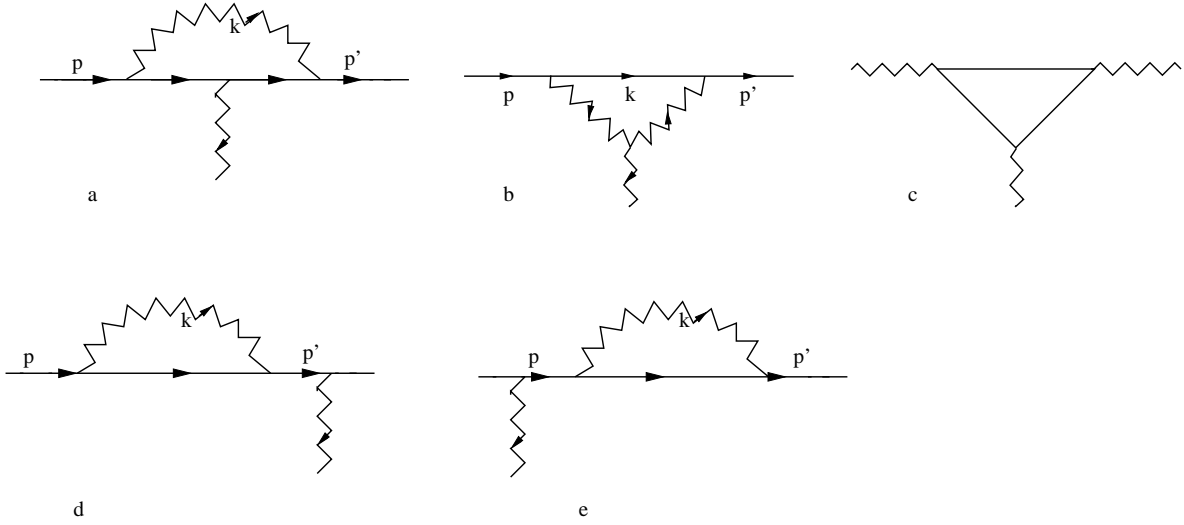}} \caption{{\it Vertices
correction.}} \label{fig4.5}
\end{figure}

The 1-loop corrections to the CS-matter field vertex are given by the
graphs in Figs.  \ref{fig4.5}(a)and \ref{fig4.5}(b).  The on-shell analytic
expression associated to the graph \ref{fig4.5}(a) is

\begin{equation}
\overline u(p^\prime)\Gamma^{a,\mu}_{\mbox {a}}u(p)=
\frac{g^3}{\Theta}[T^b T^a T_b] \int \frac{d^3k}{(2\pi)^3}
\frac{\varepsilon_{\rho\sigma\lambda} {\bar k}^\lambda \bar u(p^\prime)[
\gamma^\sigma(\not \! p^\prime- \not \! k+ m)\gamma^\mu(\not \! p- \not \! 
k+ m)\gamma^\rho]u(p)}{[(p-k)^2-m^2 + i \epsilon][(p^\prime-k)^2-m^2 + 
i \epsilon][-{\bf k}^2]}.
\end{equation}
Here and in what follows the isospin indices $(n,m)$ will omitted.
Up to the group factor $T^b T^a T_b$, this agrees with the vertex
for the Abelian theory discussed in \cite{Marcelo1}. Using dimensional
regularization, the result can be read from that reference but for general
momenta it is not particularly illuminating. Nevertheless, for small
momenta (i. e., for $|{\bf p}|\approx |{\bf p^\prime}|\ll m$) a great
simplification occurs and one finds ($\eta =\frac{|{\bf p}|}{m}$) 

\begin{equation}
\overline u(p^\prime)\Gamma_{\mbox{a}}^{a,0}u(p) = {\cal O} ( \eta^2 )\; ,
\label{4.51} \\
\end{equation}

\begin{equation}
\overline u(p^\prime)\Gamma_{\mbox{a}}^{a,i}u(p) =\frac{g^3}{4\pi\Theta} [T^b T^{a}T_b]
\frac{1}{2m} [P^{i} - i \varepsilon^{ij} q_j] + {\cal O} (\eta^2)
\; , \label{4.55}
\end{equation}

\noindent
where $P^i = p^i + p^{\prime i}$ and $q=p^{\prime i} -p^i$

Similarly, the graph \ref{fig4.5}(b) which corresponds to

\begin{equation}
\overline u(p^\prime)\Gamma^{a,\mu}_{\mbox {b}}u(p)=-
\frac{g^3}{\Theta}[ \varepsilon^{abc}T_b 
T_c]\int \frac{d^3k}{(2\pi)^3}\frac{\bar u(p^\prime)[
\gamma^\sigma(\not \! k+ m)\gamma^\beta] u(p)\varepsilon^{\mu\sigma\rho}
\varepsilon_{\sigma \beta \lambda} \overline {(p-k)^\lambda} 
\varepsilon_{\alpha
\rho\xi} \overline {(p^\prime-k)^\xi}}{[k^2-m^2+i\epsilon]({\bf p} - 
{\bf k})^2
({\bf p}^\prime - {\bf k})^2}
\end{equation}

\noindent
gives for small momenta the result

\begin{equation}
\overline u(p^\prime)\Gamma^{a,0}_{\mbox {b}}u(p) = {\cal O} (\eta^2)
\end{equation}

\noindent
and

\begin{eqnarray}
\overline u(p^\prime)\Gamma^{a,i}_{\mbox {b}}u(p) = \frac{g^3}{8\pi m\Theta} [
\varepsilon^{abc} T_c T_b] \left\{ P^{i} + i \varepsilon^{ij} q_j
\left[ 1 + \log \left(\frac{4m^2}{{\bf q}^2} \right) \right]
\right\}. \label{4.56a1}
\end{eqnarray}

The renormalized vertex part is defined by $\Gamma^{a\mu}_{R}= Z_1
\Gamma^{a,\mu}$ where $\Gamma^{a\mu} = -g T^a\gamma^\mu +
\Lambda^{a\mu}$ is the unrenormalized one. Fixing the vertex
renormalization constant $Z_1$ by the condition that for ${\bf p}={\bf
p}^\prime=0$ and $p^0=p^{\prime 0}=m$

\begin{equation}
\bar u \Gamma^{a\mu}_{R} u = -g T^a g^{0\mu}
\end{equation}

\noindent
we get $Z_1=1$, so that up to 1-loop there is no coupling constant
renormalization.  This result is also in accord with the computation
of the correction to the trilinear CS vertex shown in Fig
\ref{fig4.5}(c); simple symmetry considerations shown that the result is
finite and no counterterm is necessary.  Actually, the graph
\ref{fig4.5}(c) plus the graphs with four external gauge lines and the
polarization tensor give an induced Yang-Mills term (and also a finite
correction for the Chern-Simons term), as commented before. However,
up to one-loop the graph \ref{fig4.5}(c) does not contribute to the
scattering and for that reason it will not be considered any longer.

Summarizing, up to one-loop one needs just a mass renormalization
counterterm to fix the fermion mass. There are neither vertex nor wave
function renormalizations.

Although not 1PI, we have drawn in Fig.\ref{fig4.5} the graphs
\ref{fig4.5}(d) and \ref{fig4.5}(e) which are needed to compute the
anomalous magnetic moment of the fermions. At low momenta these graphs
give the contributions

\begin{eqnarray}
\overline u\Gamma^{a,0}_{\mbox{d}}u &=& \overline u\Gamma^{a,0}_{\mbox{e}}u 
= {\cal O}(\eta^2) \label{4.56b}\\ 
\overline u \Gamma^{a,i}_{\mbox{de}}u &=&
\overline u\Gamma^{a,i}_{\mbox{d}}u+\overline u \Gamma^{a,i}_{\mbox{e}}u
=-\frac{g^3}{4\pi\Theta} [(T^bT_b)T^{a}] \left\{ \frac{1}{2m}
[P^{i} + i \varepsilon^{ij} q_j]\right\}\; . \label{4.56c}
\end{eqnarray}

In the Abelian situation the contribution in Eq. (\ref{4.56a1}) is
absent and, in the expressions corresponding to Eqs.(\ref{4.55}) and
(\ref{4.56c}) the $P^i$ dependent part is exactly canceled. Here, due
to the group factors, to get cancellation it is necessary to take into
account the new contribution arising from (\ref{4.56a1}).  This can be
easily verified using the identity $T^b T^{a} T_b = T^{a}(T^bT_b) +
\varepsilon^{abc} T_c T_b$.  The remaining local parts occurring in
$\Gamma^{a \mu}_{\mbox{a-e}}$ will contribute to the (matrix) magnetic moment
and we get

\begin{eqnarray}
\mu_{1-loop}^{a} = \frac{ig^3}{4\pi m\Theta}[T^{a} (T^bT_b)] \; .
\label{4.57cb}
\end{eqnarray}

\noindent
This expression only differs from the corresponding result in the
Abelian case by the group factor. In the ``Abelian limit''
($g=e\sqrt{2}$ e $T= i/\sqrt{2}$) the result of \cite{Marcelo1} is
recovered.

\begin{figure}
  \centerline{ \epsfbox{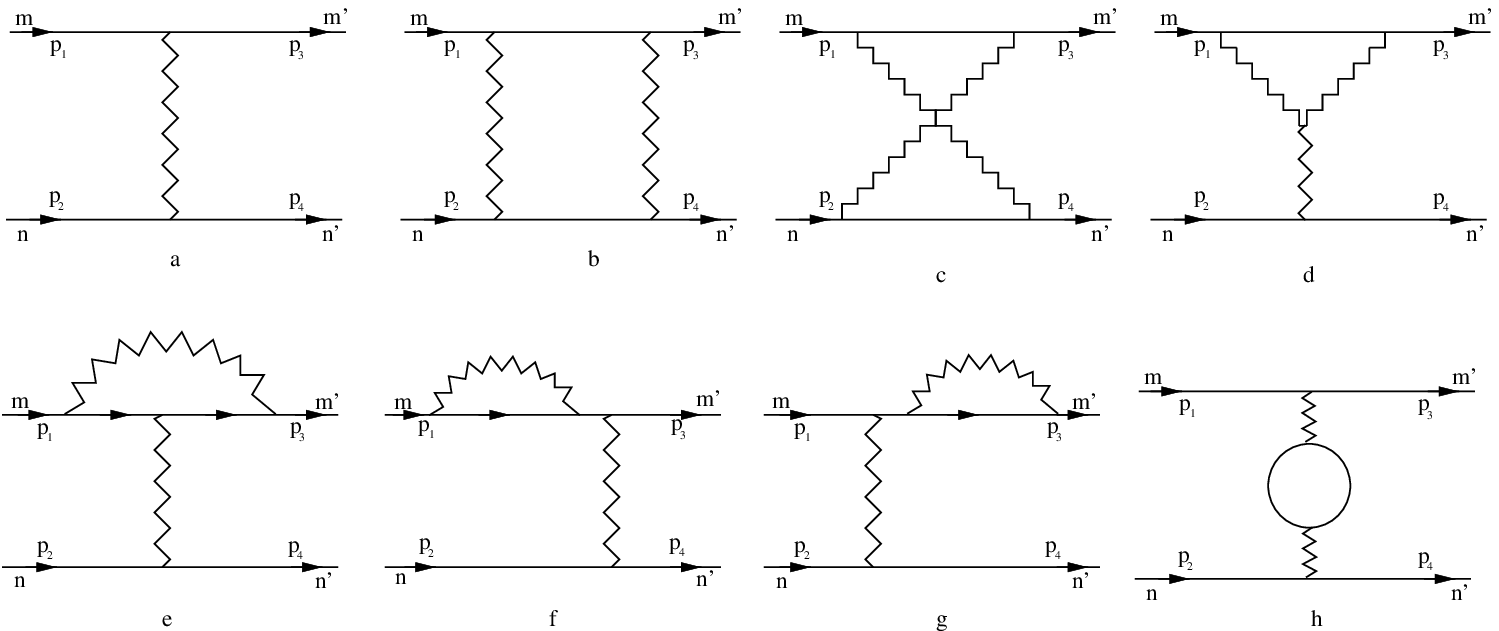}} \caption{\it {Fermion-Fermion
      scattering - relativistic theory}. Similar graphs in which
    self-energies or vertex parts are inserted in the bottom lines
    have not been drawn, for convenience.} \label{fig4.6}
\end{figure}

To complete our discussion of the one-loop properties of the model one
still has to calculate the fermion-fermion
scattering. Fig. \ref{fig4.6} shows the contributing graphs. The only
tree level graph, depicted in Fig. \ref{fig4.6}(a), furnishes

\begin{eqnarray}
{\cal M}_{\mbox{a}} (\theta) = [\bar{u}({\bf
p_4})\Gamma^{b,\nu}(p_2,p_4) u({\bf p_2})] D_{\nu\mu}^{ba} (q)
[\bar{u}({\bf p_3})\Gamma^{a,\mu}(p_1,p_3) u({\bf p_1})].
\label{4.57}
\end{eqnarray}

\noindent
To the leading order of ${\bf p}/m$,  this gives
\begin{eqnarray}
{\cal M}_{\mbox{a}} (\theta) = \frac{ig^2}{m\Theta}[T^{a}\otimes
T_a] \left\{ 1 + i \frac{\sin\theta}{(1-\cos\theta)} \right\}\; ,
\label{4.58}
\end{eqnarray}
which  exactly agrees with that obtained previously for the
nonrelativistic theory.

The one-loop graphs are represented in 
Figs. \ref{fig4.6}(b)-\ref{fig4.6}(h). 
To facilitate our computation
 we will
use an intermediate cutoff $\Lambda$, satisfying $|{\bf p}|
<<\Lambda<<m$, which separates the loop integrals in two regions. In
the {\it low} $(L)$ region $(0 \leq |{\bf k}|^2 \leq \Lambda^2)$ the
integrand is expanded in power of $1/m$, and in {\it high} $(H)$ region
$(|{\bf k}|^2 \geq \Lambda^2)$ we make a Taylor series of the integrand 
around $ |{\bf p}| \approx 0$. We will retain terms up to order 
$\eta = \frac{|{\bf p}|}m\approx \left (\frac\Lambda m\right )^2\approx 
\left ( \frac{|{\bf p}|}\Lambda\right)^2$.

 Using that

\begin{eqnarray}
S(p)= i \left[\frac{u({\bf p}) \bar{u}({\bf p})}{p^0-w_p +
i\epsilon} +\frac{v(-{\bf p}) \bar{v}(-{\bf p})}{p^0+w_p -
i\epsilon}\right] \; , \label{4.59}
\end{eqnarray}

\noindent
we may decompose the amplitude for the graph in Fig. \ref{fig4.6}(b)

\begin{eqnarray}
{\cal M}_{\mbox{b}} (\theta) &= &\int \frac{d^3 k}{(2\pi)^3}[\overline u({\bf
p}_4) \Gamma^{c,\alpha}(t,p_4)S(t) \Gamma^{d,\beta}(p_2,t)u({\bf p}_2)] 
D^{ca}_{\alpha\mu}(l^\prime)\nonumber \\
&\phantom a& [\overline u({\bf
p}_3) \Gamma^{a,\mu}(r,p_3)S(r) \Gamma^{b,\nu}(p_1,r)u({\bf p}_2)] 
D^{bd}_{\nu\beta}(l)
\end{eqnarray}

\noindent
where $l=(k^0,{\bf k}- {\bf p}_1)$, $l^\prime=(k^0,{\bf k}- {\bf p}_3)$,
$r=(w_p+ k^0,{\bf k})$ and $t=(w_p+ k^0,-{\bf k})$, into
a sum of terms

\begin{equation}
{\cal M}^{uu}_{b}+{\cal M}^{vv}_{b}
\end{equation}

\noindent
where ${\cal M}^{uu}_{b}$ and ${\cal M}^{vv}_{b}$ designate the
contributions of the $u$ and $v$ fermion wave functions to the two
internal lines of the graph. The mixed contributions in which one has
$u$ in one line and $v$ in the other vanish. After integrating in $k^0$
we obtain

\begin{equation}
{\cal M}^{uu}_{b}= \frac{ig^4}{2}[T^aT^b\otimes T_aT_b] \int 
\frac{d^2k}{(2\pi)^2} \left [ \frac{T(k,p_1) T^*(k,p_3)}{w_k-w_p}\right]
\end{equation}

\noindent
and

\begin{equation}
{\cal M}^{vv}_{b}= \frac{ig^4}{2}[T^aT^b\otimes T_aT_b] \int 
\frac{d^2k}{(2\pi)^2 }\left [ \frac{H(p_3,k) H^*(p,k)}{w_k+w_p}\right],
\end{equation}

\noindent
where 

\begin{eqnarray}
T(k,p) &=& [\overline u({\bf k})\gamma^\nu u({\bf p})]D_{\nu\beta}(k-p)
[\overline u(-{\bf k})\gamma^\nu u(-{\bf p})],\\
H(p,k) &=& [\overline u({\bf p})\gamma^\nu v(-{\bf k})]D_{\nu\beta}(k-p)
[\overline u(-{\bf p})\gamma^\nu v({\bf k})].
\end{eqnarray}

Introducing the intermediate cutoff to separate  the {\it low} and
{\it high} parts we get

\begin{eqnarray}
{\cal M}_{\mbox{b}Low}^{uu} (\theta) &=&\frac{ig^4}{4\pi m\Theta^2}
[T^{a} T^b \otimes T_a T_b] \left\{ \log
\left(\frac{\Lambda^2}{{\bf  q}^2} \right) + {\cal O}(\eta)
\right\} \label{4.68},\\ 
{\cal M}_{\mbox{b}High}^{uu} (\theta)
&=&\frac{ig^4}{4\pi m\Theta^2} [T^{a} T^b \otimes T_a T_b] \left\{
\log \left(\frac{2 m^2}{\Lambda^2} \right) + {\cal O}(\eta)
\right\} \label{4.69},\\ 
{\cal M}_{\mbox{b}Low}^{vv} (\theta)
&=&\frac{ig^4}{4\pi m\Theta^2} [T^{a} T^b \otimes T_a T_b] \left\{
{\cal O}(\eta) \right\} \label{4.70},\\
 {\cal M}_{\mbox{b}High}^{vv}
(\theta) &=&\frac{ig^4}{4\pi m\Theta^2} [T^{a} T^b \otimes T_a
T_b] \left\{ \log \left( 2  \right) + {\cal O}(\eta) \right\}\; .
\label{4.71}
\end{eqnarray}

\noindent
Putting these results together we arrive at

\begin{eqnarray}
{\cal M}_{\mbox{b}} (\theta)=\frac{ig^4}{4\pi m\Theta^2} [T^{a}
T^b \otimes T_a T_b] \left\{ \log \left(\frac{4m^2}{{\bf  q}^2}
\right)  \right\}. \label{4.71a}
\end{eqnarray}

\noindent
as the leading contribution.

For the crisscross graph, Fig. \ref{fig4.6}(c), we proceed analogously and
obtain (in this case what survives are the mixed  $uv$ and $vu$ contributions)

\begin{eqnarray}
{\cal M}_{\mbox{c}Low} (\theta) &=& -\frac{ig^4}{2\pi m\Theta^2}
 [T^b T^{a} \otimes T_a T_b]
\left\{ \frac12 \log \left( \frac{\Lambda^2}{{\bf q}^2}\right) +
{\cal O} (\eta) \right\} \label{4.74} ,\\ 
{\cal M}_{\mbox{c}High}
(\theta) &=& \frac{ig^4}{2\pi m\Theta^2}
 [T^b T^{a} \otimes T_a T_b]
\left\{  1+\frac12 \log \left( \frac{\Lambda^2}{4m^2}\right) +
{\cal O} (\eta) \right\} \; . \label{4.75}
\end{eqnarray}

\noindent
i.e.,

\begin{eqnarray}
{\cal M}_{\mbox{c}} (\theta) = \frac{ig^4}{2\pi m\Theta^2}
 [T^b T^{a} \otimes T_a T_b]
\left\{1+ \frac12 \log \left( \frac{{\bf q}^2}{4m^2}\right)
 +{\cal O} (\eta) \right\}\; .
\label{4.75a}
\end{eqnarray}

The
graph \ref{fig4.6}(d) does not exist in the Abelian theory but here it is
essential to cancel the extra contribution coming through
group factors in other graphs. It corresponds to

\begin{eqnarray}
M_d &=& \int \frac{d^3k}{(2\pi)^3}\left \{[\overline u(p_4) \Gamma^{b,\nu}
u(p_2)]\,\,D^{ba}_{\nu\mu}(q)\,\,\Gamma^{ac'd',\mu\sigma\rho}
\,\, D^{dd'}_{\sigma\alpha} (k -p_3)\right .\nonumber \\
&\phantom {a} &\left. D^{cc'}_{\beta\rho}(k-p_1)\,\,[\overline u(p_3)\, 
\Gamma^{d,\alpha}\,S(k)\,
\Gamma^{c,\beta} u(p_1)]\right \}
\end{eqnarray}

\noindent
has a {\it low}
and {\it high} momentum parts given by

\begin{eqnarray}
M_{\mbox{d}Low}&=&\frac{ig^4}{8 \pi m\Theta^2} [ 
\varepsilon^{acd} T_a \otimes T_d T_c] \left\{\frac{ }{ } 1
+i\frac{\sin\theta}{1-\cos \theta} \right. \nonumber \\ &+& \left.
\left[ \log \left(\frac{\Lambda^2}{{\bf q}^2} \right)
-\frac{\Lambda^2}{2 m^2} - (1+2\cos \theta ) \frac{{\bf
p}^2}{\Lambda^2} \right] \right\}\; ,\nonumber \\ \label{4.81}
\end{eqnarray}

\noindent
and 

\begin{eqnarray}
M_{\mbox{d}High}=\frac{ig^4}{8 \pi m\Theta^2} [ 
\varepsilon^{acd} T_a \otimes T_d T_c] \left\{-\log
\left(\frac{\Lambda^2}{4m^2} \right) +\frac{\Lambda^2}{2 m^2} +
(1+2\cos \theta ) \frac{{\bf p}^2}{\Lambda^2} \right\}\; ,
\label{4.82}
\end{eqnarray}

\noindent
respectively.
Summing Eqs. (\ref{4.81}) and (\ref{4.82}) we get

\begin{eqnarray}
M_{\mbox{d}}=\frac{ig^4}{8 \pi m\Theta^2} [  \varepsilon^{acd}
T_a \otimes T_d T_c] \left\{ 1 +i\frac{\sin\theta}{1-\cos \theta}
+ \log \left(\frac{4m^2}{{\bf q}^2} \right) \right\}\; .
\label{4.83}
\end{eqnarray}

Finally, incorporating the radiative correction, Figs.
\ref{fig4.6}(e)-\ref{fig4.6}(h), we obtain

\begin{eqnarray}
{\cal M}_{\mbox{e-g}} (\theta) = &-& \frac{ig^4}{4\pi m\Theta^2}
[T^{a}\otimes (T^bT_b) T_a]  \nonumber\\
&-&\frac{ig^4}{8\pi m\Theta^2} [ \varepsilon^{abc}T_{a}\otimes
T_c T_b] \left\{ 1
    +i\frac{\sin\theta}{(1-\cos\theta)}
 \right\}\; .
\label{4.84}
\end{eqnarray}
\begin{eqnarray}
{\cal M}_{\mbox{h}} (\theta) =\frac{ig^4}{24 \pi m\Theta^2}
[T^{a}\otimes T_a]\; . \label{4.85}
\end{eqnarray}

Summing all these contributions and using the relation (\ref{1.3})
to simplify the result, we get  the total one-loop amplitude

\begin{eqnarray} {\cal M}_{1-loop} (\theta) = \frac{ig^4}{4\pi
m\Theta^2} \left\{ \frac{3}{8} [{\it I}\otimes {\it I} ]+
\frac{2}{3}  [T^{a}\otimes T_a]\right\}\; . \label{4.87}
\end{eqnarray}

\section{Conclusions}

In this work we studied the scattering of isospin 1/2 fermionic
particles interacting through a non-Abelian Chern-Simons field. In the
nonrelativistic formulation we found that, up to a finite constant term,
there is no one-loop correction to the tree approximation to the
scattering amplitude. This is similar to what happens in the scalar
theory where the constant one-loop contribution may be eliminated by a
finite quartic counterterm \cite{Malacarne}. 

We have also considered the same problem starting from the fully
relativistic theory. After discussing the one-loop renormalizability
of the model and determining anomalous contributions to the matrix
magnetic moment of the fermions, we considered the low momenta limit
of the two body scattering amplitude obtaining a nonvanishing one-loop
contribution.  This result, shown in Eq. (\ref{4.87}), is a correction
to the scattering which does not appear in the nonrelativistic
theory. It is a leading order contribution and implies that the
effective low momentum Lagrangian contains a four-fermion self
interaction with a coupling which can be read from Eq. (\ref{4.87}).
These terms can not be eliminated by adding counterterms to the
original Lagrangian (\ref{4.22}) without destroying the
renormalizability of the relativistic model.  Furthermore, as also
happens in the Abelian case, these new terms come from the {\it high}
part of the original theory and could not be suspected in a direct
nonrelativistic approach.


\begin{thebibliography}{99}
\bibitem{Lee} T. Lee and  P. Oh, Ann. Phys. (N.Y.) {\bf 235},  413 (1994); 
H. Lo and J. Preskill, Phys. Rev.D  {\bf 48}, 4821 (1993).
\bibitem{WuYang} T. T. Wu and C. N. Yang, Phys. Rev. D {\bf 12}, 3845 (1975).
\bibitem{Horvathy} P. A. Horvathy, Phys. Rev. D {\bf 33},  07 (1986).
\bibitem{Verlinde} E. Verlinde, {\it Proceedings of the International
  Collloqium on Modern Quantum Field Theory}, Bombay, January 1990.
\bibitem{Witten} W. Witten, Nucl. Phys. {\bf B311},  46 (1988).
\bibitem{Alford} M. G. Alford and F. Wilczek, Phys. Rev. Lett. {\bf
  62},  1071 (1989); Ph. de Souza Gerbert, Phys. Rev. D {\bf 40}, 1346 (1989).
\bibitem{Wilczek} F. Wilczek and Y. S. Wu, Phys. Rev. Lett. {\bf
  65},  13 (1990).
\bibitem{Alford2} M. G. Alford, J. March-Russel and F. Wilczek,
  Nucl. Phys.  {\bf B337},  695 (1990).
\bibitem{Kibble} T. W. B. Kibble, J. Phys {\bf A9},  1387 (1976).
\bibitem{Vilenkin} A. Vilenkin, Phys. Rev. D {\bf 23},  852 (1981).
\bibitem{Aharonov} Y. Aharonov and D. Bohm, Phys. Rev. {\bf 115}, 485 (1959).
\bibitem{Rafaeli} E. L. Feinberg, Sov. Phys. Usp. {\bf 5},  753 (1963);
E. Corinaldesi and F. Rafeli, Am. J. Phys. {\bf 46},  1185 (1978);
K. M. Purcell and W. C. Henneberger,  Am. J. Phys. {\bf 46}, 1255 (1978).
\bibitem{Bergman} O. Bergman and G. Lozano, Ann. Phys. (N.Y.) {\bf 229},
 416 (1994).
\bibitem{Malacarne} M. Gomes, L. C. Malacarne and A. J. da Silva, Phys. Rev. D
{\bf 59}, 045015 (1999).
\bibitem{Bak} D. Bak and O. Bergman, Phys. Rev. D {\bf 51}, 1994 (1995).
\bibitem{Jorge} M. Gomes, J. M. C. Malbouisson and  A. J. da Silva,
  Mod. Phys Lett. {\bf A11},  2825 (1996); Phys. Lett. {\bf A207}, 
  373 (1997).
\bibitem{Hagen1}C. R. Hagen, Phys. Rev. Lett. {\bf 64},  503 (1990);
  C. R. Hagen, Phys. Rev. D {\bf 41},  2015 (1990);
  C. R. Hagen, Phys. Rev. D {\bf 48},  5935 (1993).
\bibitem{Marcelo1}M. Fleck, A. Foerster, H. O. Girotti, M. Gomes, J. R. S. 
Nascimento and
  A. J. da Silva, Int. J. Mod. Phys. A {\bf 12} 2889 (1997; H. O. Girotti, M. 
Gomes, J. R. S. Nascimento and
  A. J. da Silva, Phys. Rev. D {\bf 56},  3623 (1997)
\bibitem{Marcelo3} M. Gomes, L. C. Malacarne and A. J. da Silva, Phys. Rev. D
{\bf 60}, 125016 (1999).
\bibitem{Marcelo2} M. Gomes and A. J. da Silva, Phys. Rev. D {\bf 57},
   3579 (1998).
\bibitem{Ferrari} F. Ferrari and L. Lazzizzera,
Int. J. Mod. Phys. {\bf A13}, 1773 (1998);
\bibitem {notacao}  For the $\gamma$-matrices we adopt the representation
$\gamma^{0}=\sigma^{3},\gamma^{1}=i\sigma^{1},\gamma^{2}=i\sigma^{2}$,
where $\sigma^{i}, i=1,2,3,$ are the Pauli spin matrices. The
positive and negative energy solutions of the free Dirac equation are
given by

\[
u(\vec{p})\,\,=\,\, \left( \frac{w_p+m}{2\,w_p}
\right)^{\frac {1}
{2}}\,\left[ \begin{array}{c} 1 \\
\frac{p^{2}\,-\,ip^{1}}{w_p\,+\,m} \end{array} \right] \,\,\quad 
v(\vec{p})\,\,=\,\, \left( \frac{w_p+m}{2\,w_p}
\right)^{\frac {1}
{2}}\,\left[ \begin{array}{c} \frac{p^{2}+i p^{1}}{w_p\,+\,m} \\
1\end{array} \right] \,\,.
\]
where $w_p=(m^2+{\vec p}^2)^{1/2}$ and the normalizations were chosen so that
$\bar u u = - \bar v v= m/w_p$.

\bibitem{Adkins} G. S. Adkins, Phys. Rev. D {\bf 27},  1814 (1983).

\end{thebibliography}
\end{document}